\documentclass{article}

\usepackage{PRIMEarxiv}

\usepackage[utf8]{inputenc} % allow utf-8 input
\usepackage[T1]{fontenc}    % use 8-bit T1 fonts
\usepackage{hyperref}       % hyperlinks
\usepackage{url}            % simple URL typesetting
\usepackage{booktabs}       % professional-quality tables
\usepackage{amsfonts}       % blackboard math symbols
\usepackage{nicefrac}       % compact symbols for 1/2, etc.
\usepackage{microtype}      % microtypography
\usepackage{lipsum}
\usepackage{fancyhdr}       % header
\usepackage{graphicx} % graphics
\usepackage{float}
\usepackage{subfig}
\usepackage{comment}
\usepackage[round]{natbib}
\graphicspath{{media/}}     % organize your images and other figures under media/ folder

\title{Numerical Approximation Methods for Antenna Radiation Patterns for Motus Wildlife Tracking Systems
}

\author{
  Erik Carlson, Douglas Gobeille \\
  Department of Physics \\
  University of Rhode Island \\
  Kingston, RI\\
  \texttt{\{erikvcarlson, dgobeille\}@uri.edu} \\
   \And
  Robert Deluca \\
  Physics Systems LLC \\
  Xenia, OH\\
  \texttt{robertdeluca@mac.com} \\
    \AND
  Pam Loring \\
  Division of Migratory Birds \\
  US Fish and Wildlife Service \\
  Charlestown, RI\\
  \texttt{pamela\_loring@fws.gov}
}

\begin{document}
\maketitle
\begin{abstract}
As plans for offshore wind energy development increase in the US, the developing methods to monitor migratory birds and bats offshore is an important area of research. To contribute to this research, current guidance recommends the deployment of Motus Wildlife Tracking stations by each developer for pre- and post- construction monitoring. To understand the characteristics of each of the stations, calibration techniques are recommended for each deployment. Despite this, previous attempts to calibrate these stations has failed to provide sufficient detail to allow for high-resolution tracking techniques. In this paper, we introduce an affordable and robust methodology to calibrate stations in the off-shore environment and develop processes to turn this raw calibration data into a numerical approximation of the antennas radiation pattern. 
\end{abstract}

% keywords can be removed

\section{Introduction}
With the increase in off-shore wind (OSW) lease areas in the last several years, monitoring offshore bird and bat species has become an issue of upmost importance. For small aerofauna ($<150$ g), technologies traditionally utilized for monitoring individual movements utilizing GPS tracking are limited due to the nature of their weight. Following the work of \cite{loring_2016}, digital VHF transmitters (RF-Tags) were identified as a safe alternative for monitoring small-bodied species including the federally threatened Piping Plover (Charadrius melodus) and the federally endangered Roseate Tern (Sterna dougallii). This work is conducted in coordination with the Motus Wildlife Tracking System (Motus) an international automated radio  \\ \\  While these techniques were successful at tracking the movements of these aerofauna on a regional level, more precise tracking required for monitoring these aerofauna relative to OSW lease areas are required to directly monitor site specific movements in the offshore areas. To accomplish this, the intersection of near-simultaneous detections from $>3$ stations was purposed (\citealp{2018arXiv181002416B}; \citealp{2018arXiv180511171J}). While the underlining theory behind this methodology is correct, the utilization of theoretical radiation patterns fails to take into account the nuances of each stations deployed environment. For example, as the station ages and dielectric material builds up on the antenna, the radiation pattern of each deployed antenna will change dramatically from the theoretical pattern. Additionally, the theoretical pattern does not account for site-specific factors including antenna blinds spots, reflections and other interference which must be quantified numerically. \\ \\  This paper's goal, in support of upcoming federal guidance documents, is to outline the methodology and analysis of calibration data of OSW and land-based Motus Stations in an effort to quantify and standardize methods to produce numerical approximation the radiation pattern of yagi and omni-direction antennas. 

\section{Methods}
The primary method for producing the models for our radiation patterns revolves around co-locating GPS locations with RF-Tag detections. Although various types of calibration data collection have been used in previous efforts, there is a need to standardized key elements including in the GPS and tags temporal resolution, power regulation and final data analysis in order to apply the method consistently and collect the amount and quality of data needed to
estimate radiation patterns.
\subsection{Pre-Calibration Equipment} \label{sec: Pre-Cal-Equip}
To test our calibration methods, three stations were deployed and tested throughout this investigation. Each station was 20ft tall and was deployed with four horizontally mounted circularly polarized yagi-uda antennas and one vertically mounted omni-direction antenna, all operating at 434 MHz. One station was located at a local residence for initial testing and to represent a deployed land-based station. Two stations were placed at Block Island, Rhode Island at South East Lighthouse (SEL) and Black Rock Beach (BRR) to represent shore-based monitoring systems. Finally, one station was placed at the Block Island Wind Farm (BIWF) Turbine 1 to represent a
deployed station in an off-shore environment. The locations of each of these stations is located in Table \ref{tab:station_locs} \\ \\ 

\begin{table}[h!]
    \centering
    \begin{tabular}{ | c c c | }
    \hline
    Station Name & Latitude & Longitude \\ 
        \hline
    South East Lighthose & 41.1534 & -71.5521\\
    Black Rock Beach & 41.1479 & -71.5901 \\
    BIWF Turbine 2 & 41.1418 & -71.5302 \\

    \hline
    \end{tabular}
\caption{ \label{tab:station_locs} Station Locations}
\end{table}
A standard Cell Track Tech (CTT) PowerTag operating at 434 MHz was utilized for all of our calibration efforts with a one-second ping rate to correspond with the ping rate of our GPS. These tags were also outfitted with larger batteries and activators to conserve battery life in between calibrations. Utilizing a PowerTag over the CTT LifeTag ensured a constant ping strength over the course of the calibration cycle. \\ \\ To provide GPS location, we utilized a Garmin Edge 130. This device provided high frequency (1 Hz) GPS locations and relied on barometric pressure readings to provide accurate elevation information. It was also light enough to be used in all successful calibration methods. Finally, this device was not cost prohibitive for station operators to purchase and utilize in the field for their own calibration studies.  
\subsection{Traditional Calibration Methods} \label{sec:Kite_and_Stick}
Traditional calibrations methods, such as those used in \cite{2018arXiv180511171J}, attached a test tag to a partially frozen bird carcass that was suspended from a kite and flown from the back of a boat. During Janaswamy's calibration surveys, the boat was approximated to travel at a constant speed and transmitter at an approximate height of $30$ m. This approximation introduced large errors into the calibration data. Additionally, a boat hook with a test tag was attached to the hull of the boat and carried throughout their calibration process to approximate the radiation pattern at an assumed $3$ m ASL (above sea-level). \\ \\ 
To continue the theme of inexpensive, minimal effort calibration efforts, we built on and improved the techniques used in \cite{2018arXiv180511171J}. First, we utilized a non-conducting pole which lowered the interference of the tag's signal. Secondly, we adhered a barometric GPS to our kite which allowed for precise measurement of the tag's altitude and position.   

\subsection{Small Unmanned Aerial System Operations} \label{sec:UAS}
Throughout this effort EC conducted Unmanned Aerial System (UAS) operations utilizing a DJI Mavic Air 2 with a co-located GPS and RF-Tag. Flying this UAS around each of the stations provided detailed calibration information between $0$ m AGL (above ground-level) to $122$ m AGL. These flights are a far superior method for collecting detailed altitude information throughout the study as they allow for fine control of the position and altitude of the UAS. This fine control allows the user to fly the UAS precisely into areas where previous calibration data was not collected, thus optimizing the time and cost of each calibration survey. 
\subsection{Aviation Calibration} \label{sec:CAP}
To gather high ($>300$ m AGL) altitude information about each station, small aviation aircraft were utilize to conduct flights over the stations in the study. These flights, conducted by with a Cessna 182, provided similar calibration information to UAS operations. However, due to the stall speed of the Cessna 182, high resolution calibration was not possible. This was abated by utilizing the winds aloft to lower the overall ground speed of the aircraft. 

\section{Results and Analysis} \label{sec:Results}
\begin{figure}[H]
\centering
\includegraphics[width=8cm]{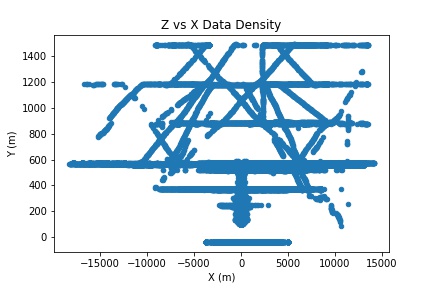}
\caption{Detected Points from South East Lighthouse in the Vertical Plane}
\label{fig:Altitude_Information}
\end{figure}

\begin{center}

\begin{figure}[H]
\centering
\begin{tabular}{cc}
\subfloat[Boat Calibration Track]{\includegraphics[width = 3.5in]{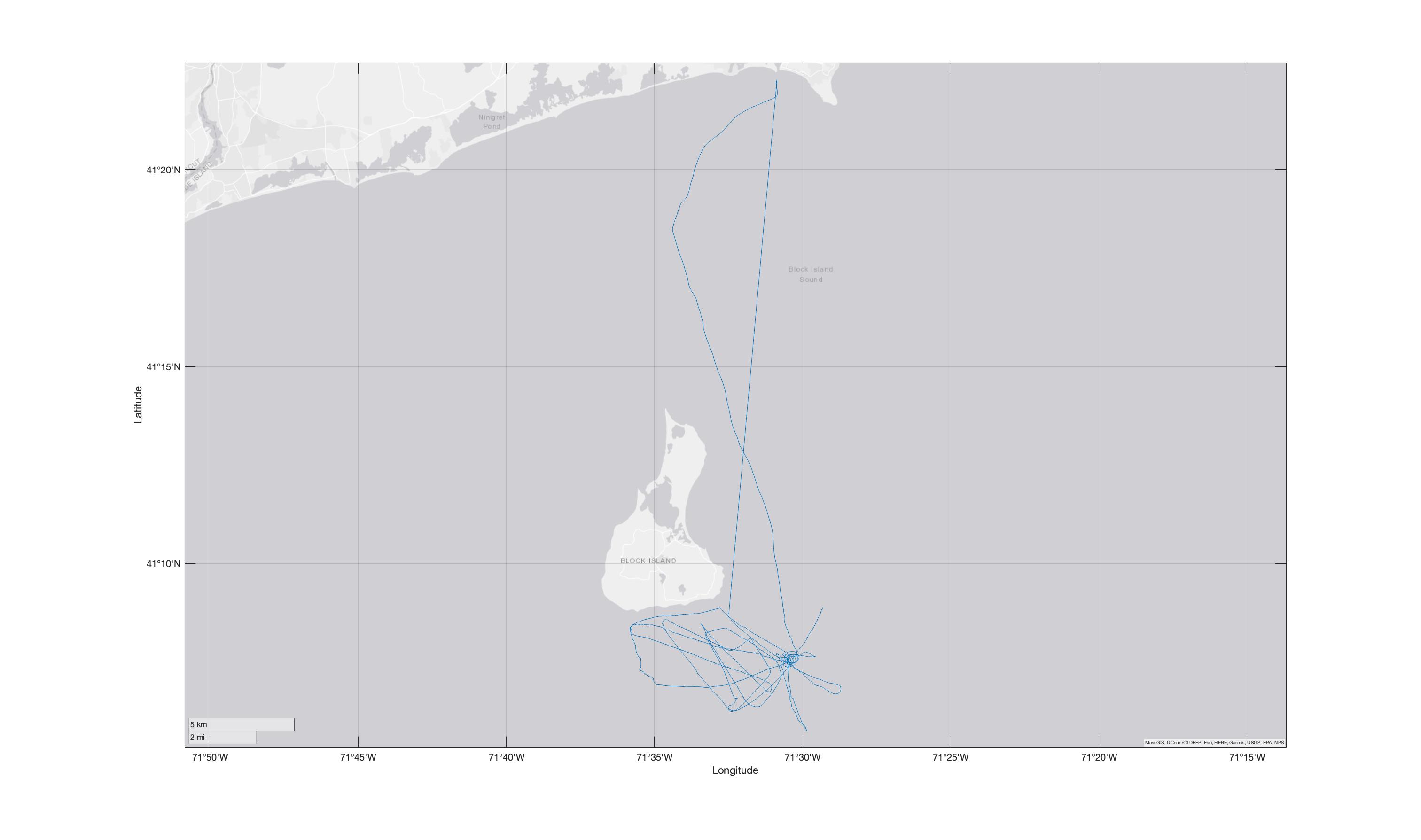}} &
\subfloat[UAS Calibration Track]{\includegraphics[width = 3.5in]{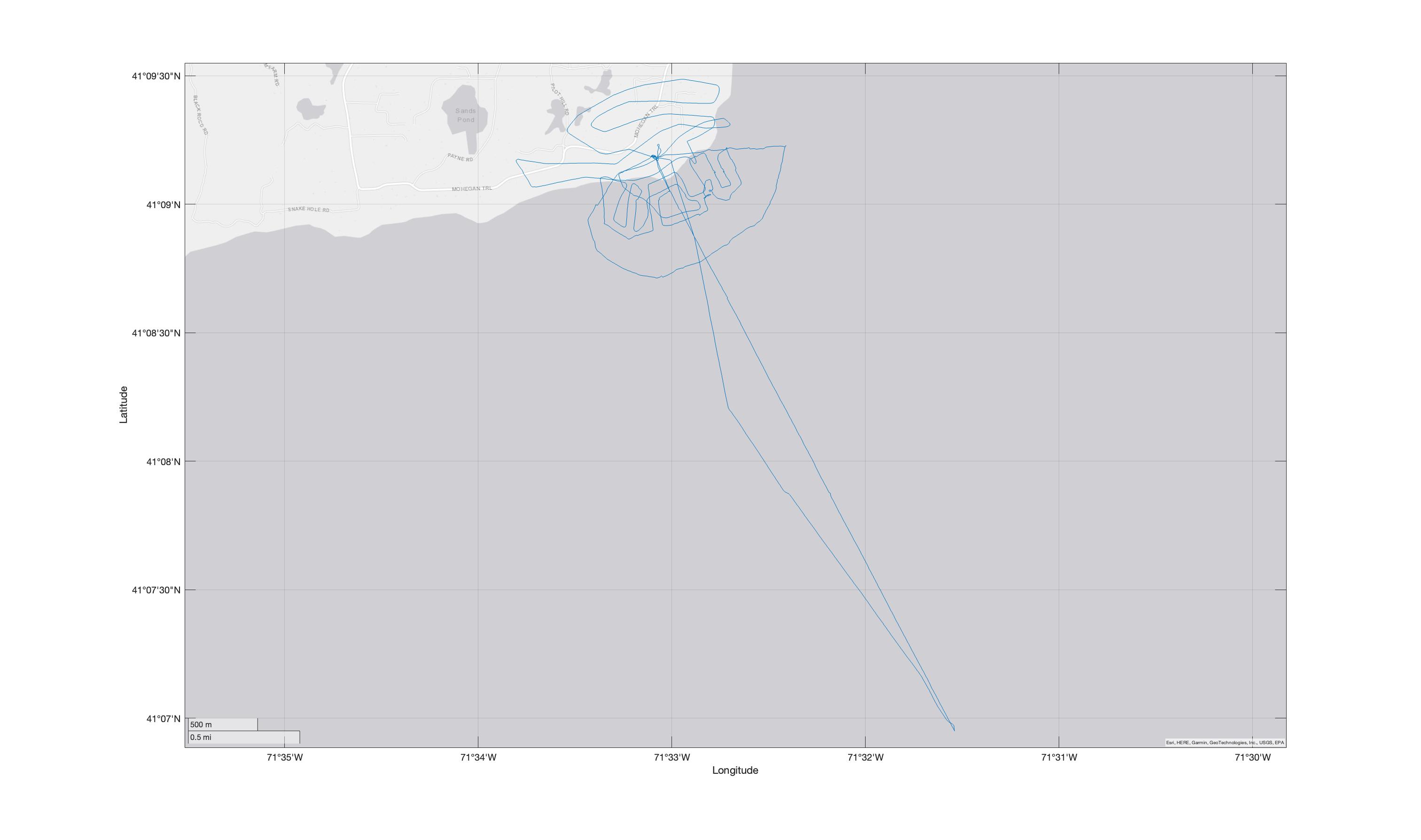}} \\
\end{tabular}
\subfloat[Aircraft Calibration Track]{\includegraphics[width = 3.5in]{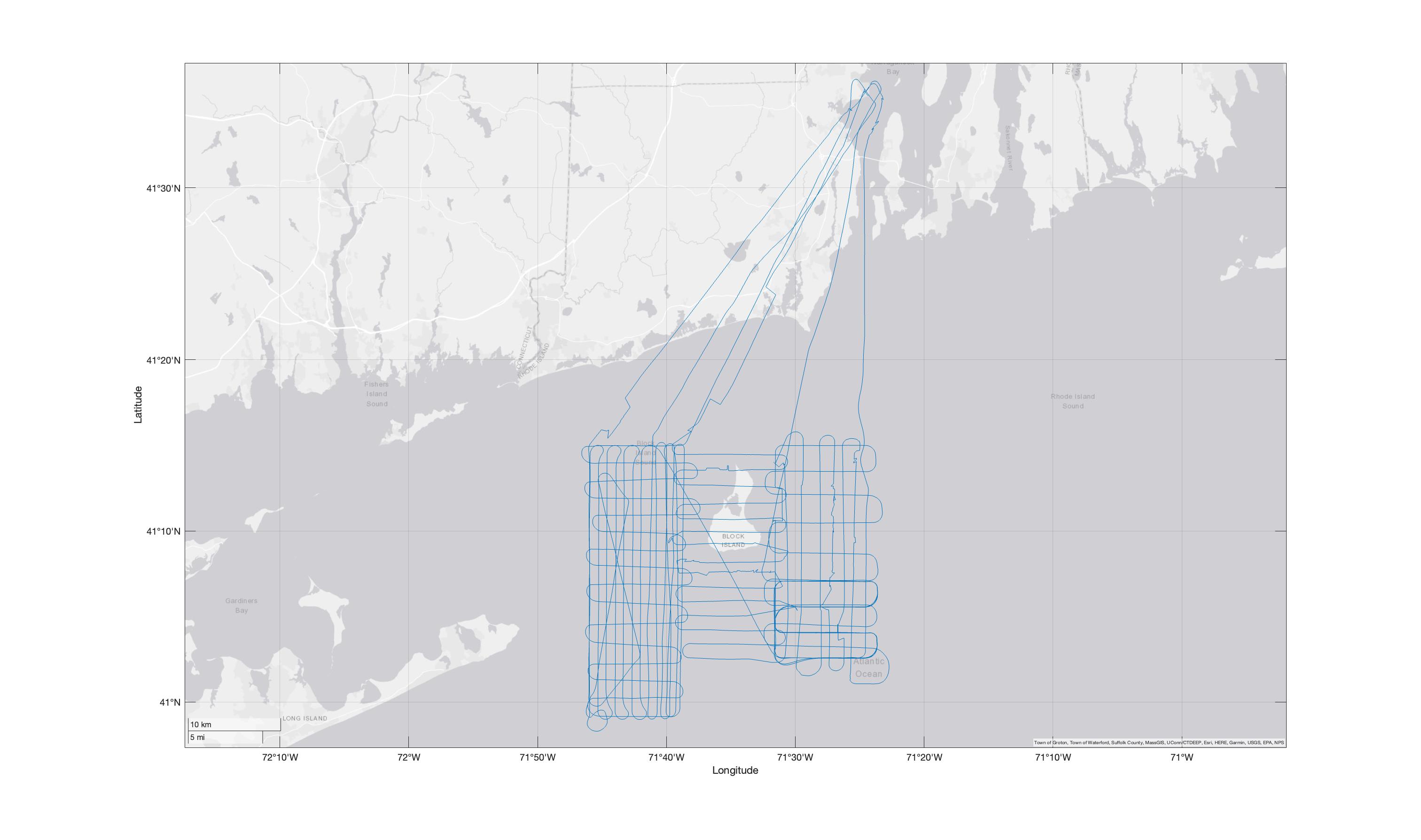}} 
\caption{}
\label{fig:boat_im1}
\centering
\begin{tabular}{cc}
\subfloat[South East Lighthouse Detections]{\includegraphics[width = 3.5in]{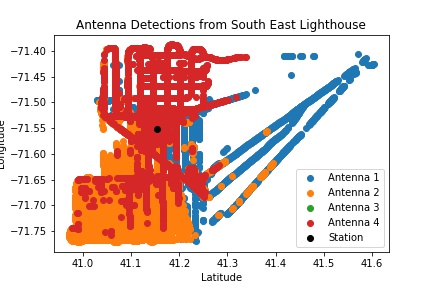}} &
\subfloat[Black Rock Detections]{\includegraphics[width = 3.5in]{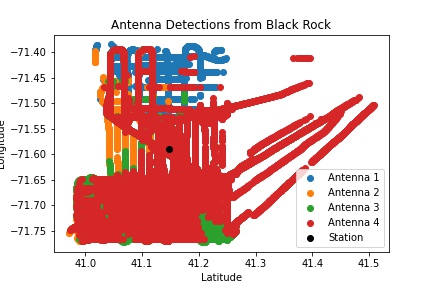}} \\
\end{tabular}
\subfloat[Turbine Detections]{\includegraphics[width = 3.5in]{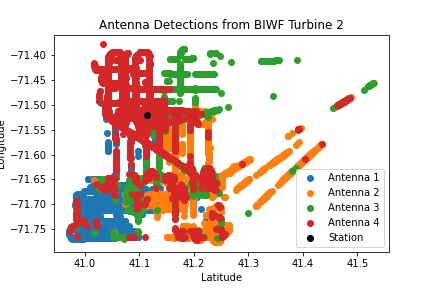}} 
\caption{}
\label{fig:SEL_XYZ}
\end{figure}
\end{center}

\begin{figure}[h]
 \centering
\begin{tabular}{c c}
\subfloat[South East Lighthouse Top-Down]{\includegraphics[width = 3.3 in]{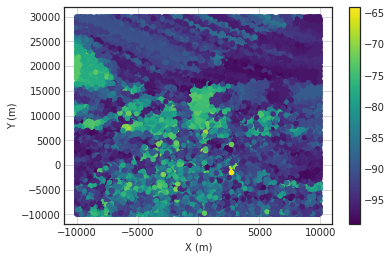}} &
\subfloat[South East Lighthouse Side-On]{\includegraphics[width = 3.3 in]{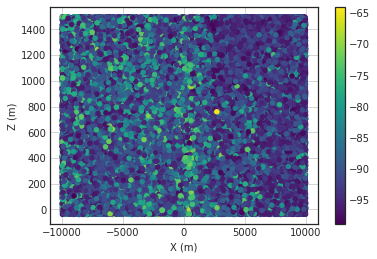}} \\
\subfloat[Black Rock Station Top-Down]{\includegraphics[width = 3.3 in]{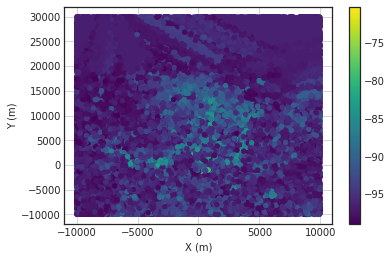}} &
\subfloat[Black Rock Station Side-On]{\includegraphics[width = 3.3 in]{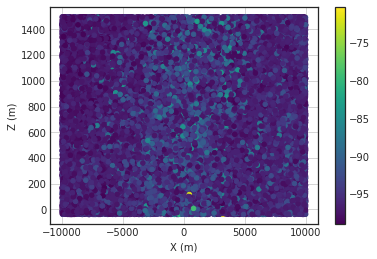}} \\
\subfloat[Turbine Station Top-Down]{\includegraphics[width = 3.3 in]{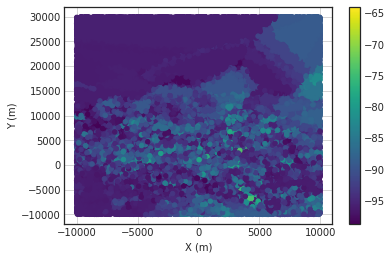}} &
\subfloat[Turbine Station Side-On]{\includegraphics[width = 3.3 in]{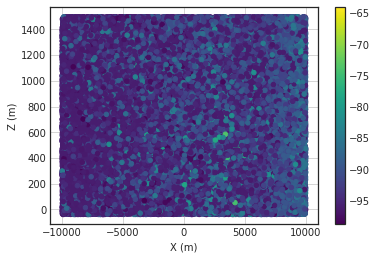}} \\
\end{tabular}
\caption{K-Means Results Colorized by dBm with Antenna Beam Pointed towards the Positive Y-Axis}
\label{fig:k_means_results}
\end{figure}

Five Calibration surveys were conducted over the course of 2021, their specific properties\footnote{Raw Calibration data is located at: https://tinyurl.com/BIWFCalData} and tracks are located in Table  \ref{tab:calibration_info} and Figure \ref{fig:boat_im1}. \\ \\
\begin{table}[H]
    \centering
    \begin{tabular}{ | c c c | }
    \hline
    Date & Calibration Type & GPS Points Collected \\ 
    \hline
    Apr 17 & UAS & 4,253 \\  
    Aug 30 & Traditional & 18,715 \\  
    Aug 31 & Traditional & 10,864 \\  
    Sep 18/19 \footnotemark & Aviation & 13,543 \\ 
    Sep 26 & Aviation & 12,277 \\
    Sep 27 & Aviation & 14,862 \\ 
    \hline
    \end{tabular}
\caption{ \label{tab:calibration_info} Calibration Event Properties}
\end{table}
\footnotetext{Calibration started on September 18 and continued past 0:00 UTC into September 19}
Upon the completion of the calibration runs, test tag data was downloaded from the Motus database for each of the land-based stations and was manually collected via USB drive from the station deployed in the Block Island Wind Farm. The tag data was first filtered to exclude all tag data from tags not used specifically for calibration. We then identified simultaneous detections of the GPS and tag, allowing us to assign a tag signal strength to a given location using the timestamps of each tag detection and GPS ping. Knowing the position of each station allowed us to transition our GPS detections to a relative position to each antenna. We present the results of this co-location for each of the three stations in Figure \ref{fig:SEL_XYZ}.
To continue our analysis of any given station, we utilized k-nearest neighbor regression to estimate a signal strength for arbitrary points around each station. To do this we generated 100,000 random points within a 20 km by 30 km by 1.5 km volume assigning values for the RSSI utilizing scikit-learn's K-neighbors regressor with 5 neighbors \citep{scikit-learn}. We present the results of this analysis in Figure \ref{fig:k_means_results}. \\ \\ 
As we can see from the results, most of our top down images have defined structure along the positive y-axis of each calibration run. This is particularly prominent in the South East Lighthouse approximation which shows the a clearly defined elliptical structure stemming from the origin and protruding to approximately 15km away. This structure aligns with theoretical approximates of the primary beam of a standard yagi-antenna beam pattern.  \\ \\ Although not obvious in the images below, each station presented some elliptical structure in line with what would be the putative primary beam of each antenna when the scaling is reduced to only include the strongest detections. We stipulate that the South East Lighthouse Station best presented this structure due to the targeted UAS calibration flights directly within the primary beam while other surveys over or under flew the primary radiation pattern resulting in an overall lesser number of strong detections. \\ \\ 
No station was able to produce any type of structure in the edge-on plane. This is due to the fact that our vertical calibration plane remained poorly sampled despite our attempts to utilize various calibration methods. We present our data density in the vertical plane in Figure \ref{fig:Altitude_Information}.
To further examine our data, we examine how well each station model reproduced the calibration data for its own station as well as the calibration data for the other stations. The results of which are presented in Table \ref{tab:error_info}.
\begin{table}[H]
    \centering
    \begin{tabular}{ | c c c c | }
    \hline
    Station 1 & Station 2 & Mean Error (dBm) & Standard Deviation (dBm) \\ 
    \hline
    Black Rock & BIWF Turbine & 5 & 3\\  
    Black Rock & Black Rock & 3 & 2\\  
    Black Rock & South East Lighthouse & 7 & 6 \\  
    South East Lighthouse & South East Lighthouse & 6 & 5 \\ 
    BIWF Turbine & South East Lighthouse & 7 & 2 \\
    BIWF Turbine & BIWF Turbine & 8 & 8\\ 
    \hline
    \end{tabular}
\caption{ \label{tab:error_info} Data Reproduction Errors}
\end{table} Of all the stations tested the Black Rock Station Model had the best results when reproducing the calibration data for its own station. This is most likely due to the absence of any large conducting materials around the station as well as clear lines of sight to the horizon. Both the South East Lighthouse and the BIWF station had large reflectors around each station causing increased error. \\ \\  Finally, we see that no two station's models reproduce each other. This is expected given the unique nature of each station's unique environment and demonstrates the need for numerical methods for calibrating Motus stations. 
\section{Conclusion}
Through this paper, we have introduced the methodology to generate numerical approximations of radiation patterns for Motus Wildlife Tracking Systems. We have demonstrated that this method can approximate the received signal strength of a tag to within an average of 8 dbm. Given the nature of decibel-milliwatts, this represents a small error for detections close to the station and a much larger error farther away from the station. To quantify this error into meters, more calibration data is needed. This methodology, in conjunction with a the analytical approach in \cite{2018arXiv180511171J} are being used to approximate the effectiveness of Motus tracking stations in Off-Shore Wind Farm application \cite{adams_sim_study}. We have also identified the time intensive nature of high-resolution calibration surveys. Future work will aim to reduce the overall survey effort required to conduct these surveys while producing more robust altitude information. This work will be used in the generation of a tracking algorithm utilizing the intersections of tessellated radiation patterns produced by this approach. 

\section*{Acknowledgments and Disclaimers}
This study was funded by the New York State Energy Research and Development Authority under Agreement No. 143771. NYSERDA has not reviewed the information contained herein, and the opinions expressed in this report do not necessarily reflect those of NYSERDA or the State of New York. Any use of trade, firm, or product names is for descriptive purposes only and does not imply endorsement by the US Government. The findings and conclusions in this article are those of the author(s) and do not necessarily represent the views of the US Fish and Wildlife Service. EC was also supported by the NASA Rhode Island Space Grant Consortium during this work. We thank Kate Williams, Evan Adams, and Andrew Gilbert at the Biodiversity Research Institute for their insights. For pilot work on Block Island, we thank Block Island Wind Farm (Orsted),
Scott Comings (The Nature Conservancy), Lisa Nolan (Southeast Lighthouse Foundation), and Brett Still (University of
Rhode Island). For calibration surveys, we thank Suzanne Paton (U.S. Fish and Wildlife Service Coastal Program), Tom
Halavik, Nathan Fueller, and Rhode Island Civil Air Patrol. 
%Bibliography
%\bibliographystyle{unsrt}  
\bibliography{references}  

\newgeometry{top=0.1in,bottom=0.1in,left=0.3in,right=0.3in}

\end{document}